\documentclass[]{svjour3}
\usepackage{graphicx}
\usepackage{amssymb}

\begin{document}

\title{The dynamics of the elliptic Hill problem : Periodic orbits and stability regions}
\titlerunning{The Elliptic Hill problem}

\author{G. Voyatzis, I. Gkolias,  H. Varvoglis}
\authorrunning{Voyatzis et al.} % if too long for running head

\institute{Department of Physics, Aristotle University of Thessaloniki\at
              Thessaloniki, Greece \\
              \email{voyatzis@auth.gr, igkoli@physics.auth.gr, varvogli@physics.auth.gr}
          }
\maketitle

\begin{abstract}
The motion of a satellite around a planet can be studied by the Hill model, which is a modification of the restricted three body problem pertaining to motion of a satellite around a planet. Although the dynamics of the circular Hill model have been extensively studied in the literature, only few results about the dynamics of the elliptic model were known up to now, namely the equations of motion and few unstable families of periodic orbits. In the present study we extend these results by computing a large set of families of periodic orbits and their linear stability and classify them according to their resonance condition.  Although most of them are unstable, we were able to find a considerable number of stable ones.  By computing appropriate maps of dynamical stability, we study the effect of the planetary eccentricity on the stability of satellite orbits. We see that, even for large values of the planetary eccentricity, regular orbits can be found in the vicinity of stable periodic orbits. The majority of irregular orbits are escape orbits. 
\end{abstract}

\section{Introduction}
The motion of a satellite around a planet, which revolves on a circular orbit around the Sun, can be modelled by the Hill's model of the three-body problem. Starting from the circular restricted three body problem (CRTBP), the Hill's approximation is achieved by translating the origin of the rotating reference frame to the planet and the unit of length is scaled by the factor $\mu^{1/3}$, where $\mu$ is the  mass parameter of CRTBP. Then we let $\mu\rightarrow 0$. The present study is restricted to the planar motion. Under these assumptions we obtain the Hill's equations  (Szebehely, 1967)
\begin{equation} \label{EqCHillEqs}
\ddot{\xi}-2\dot{\eta}=3\xi-\xi\,\rho^{-3},\quad \ddot{\eta}+2\dot{\xi}=-\eta\,\rho^{-3}
\end{equation}
where $\rho=(\xi^{2}+\eta^{2})^{1/2}$. These equations admit an integral of motion, the well known Jacobi integral
\begin{equation} \label{EqJacobi}
C_H=3\xi^2+2\rho^{-1}-(\dot \xi^2+\dot \eta^2),
\end{equation}
and are invariant under the symmetries
$$
\mathbf{\Sigma}:(t,\xi,\eta)\rightarrow (-t, \xi,-\eta)\quad \textnormal{and} \quad \mathbf{\Sigma'}:(t,\xi,\eta)\rightarrow (-t, -\xi,\eta).
$$
Although the circular Hill problem (CH) has only two degrees of freedom and is, moreover, autonomous, conservative and parameter free, it is not integrable (Meletlidou et al., 2001; Morales-Ruiz et al., 2005) and shows rich dynamics, which are depicted by computing Poincar\'{e} surfaces of section (H\'enon, 1970; Chauvineau and Mignard, 1991).

\begin{figure}
\centering
\includegraphics[width=9cm]{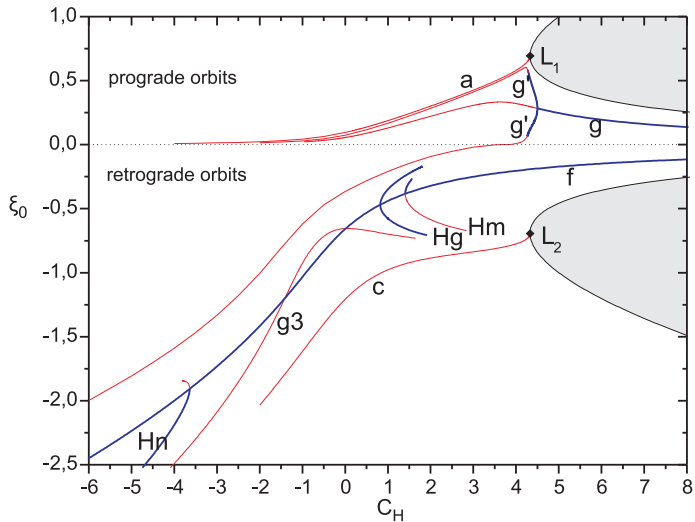}
\caption {Families of periodic orbits of the Circular Hill problem. Thick (blue) or thin (red) curves indicate stable or unstable families, respectively. Grey regions indicate forbidden areas.}
\label{FigCHfams}
\end{figure}

A detailed study of the CH problem has been given by H\'{e}non (1969,1970). He found the main families of periodic orbits, explored the phase space by using Poincar\'{e} sections and computed the width of the stability regions. As we know, the families of periodic orbits consist the backbone of the phase space. The most important of them are presented in Fig. \ref{FigCHfams} (see also H\'{e}non, 2003; Batkhin and Batkhina, 2009). The domain $\xi>0$ corresponds to prograde and $\xi<0$ to retrograde motion. The two equilibrium (Lagrange) points $L_1$ and $L_2$ are found for $C_H^L=3^{4/3}$ at the locations $\xi=\pm 3^{-1/3}$, $\eta=0$. The distance $R_H=3^{-1/3}$ defines the Hill radius. We can see from Fig. \ref{FigCHfams} that families of periodic orbits of retrograde satellites extend to distances much larger than the Hill radius and, interestingly, some of them are stable.

The Hill's approximation can be applied in the same way to the spatial problem (H\'enon, 1974).  Such a model has been used in space mission orbit design (Villac, 2003, 2008). A single averaged model for the spatial Hill's problem has been also studied, providing particular solutions (see Vashkov'yak and Teslenko (2008) and references therein). Hill models have also been derived for the case of binary satellites or asteroids (H\'enon and Petit, 1986; Chauvineau and Mignard, 1990). In these models the center of mass of the binary moves along a circular orbit around the Sun. A model where the center of the binary moves in an elliptic orbit has been given by Moons et al. (1988).  This model uses a different formalism and aims to the study of the changes of orbital elements due to close encounters.  An extensive list of references on the dynamics of satellite motion is given in Waldvogel (1999). The general three body problem may be also used for studying satellite or binary motion (Hadjidemetriou and Voyatzis, 2011). 

The elliptic Hill (EH) model can be derived by considering the same Hill's approximation assumptions used in the CH model, but letting the planet move on an elliptic orbit around the Sun. Such a model has been introduced by Ichtiaroglou (1980, 1981) who also computed a few families of periodic orbits. A further study (Ichtiaroglou and Voyatzis, 1990) showed that all periodic orbits found were strongly unstable. It is obvious that the EH is a more appropriate model than the CH for studying the dynamics of a small satellite around a planet (or an asteroid) with eccentric motion. The aim of the present work is to study the dynamics of the EH problem in more detail, focusing on the main qualitative features of the phase space and orbit evolution.

In the next section we describe briefly the EH model and discuss the existence and continuation of periodic orbits. In section \ref{SecFamilies} we present the results of the computations of families of periodic orbits, giving particular attention to the stable ones. In section 4 we explore the phase space of the model by computing various maps of stability and study the effect of the planetary eccentricity to the stability of the orbits. Our conclusions are given in section 5.

\section{The Elliptic Hill model}
Considering two primaries of masses $m_0$ (the Sun) and $m_1$ (the planet) that revolve on the plane $Oxy$ in a Keplerian ellipse around their center of mass $O$, the motion of the planet along the rotating axis $Ox$ with angular momentum $P_\theta$ (see Hadjidemetriou 1975) is described by the equation
\begin{equation} \label{EqERequP}
\ddot x_1 -Q^2 x_1^{-3}+(1-\mu)^3 x_1^{-2}=0,
\end{equation}
where $\mu=m_1/(m_0+m_1)$ and $Q=(1-\mu)P_\theta/\mu$. By assuming the initial conditions $x_1(0)=x_{10}$ and $\dot\theta(0)=1$, where $\dot\theta$ is the angular velocity, we get $Q=x_{10}^2$.

The motion of a massless body, which moves on the same plane $Oxy$, is described by the equations
\begin{equation} \label{EqERequB}
\begin{array}{l}
\ddot x=2Qx_1^{-2}\dot{y} + Q^2 x_1^{-4} x - 2Q x_1^{-3}\dot{x}_1 y+\mu (x_1-x) r_1^{-3}-\left (\mu x_1+(1-\mu)x \right ) r_2^{-3} \\
\ddot y=-2Qx_1^{-2}\dot{x} + Q^2 x_1^{-4} y + 2Q x_1^{-3}\dot{x}_1 x-\mu y r_1^{-3}-(1-\mu) y r_2^{-3}, \end{array}
\end{equation}
where $r_1^2=(x-x_1)^2+y^2$ and $r_2^2=(x+\frac{\mu}{1-\mu} x_1)^2+y^2$. 
Equations (\ref{EqERequP}) and (\ref{EqERequB}) are the equations of the {\em restricted elliptic three body problem} in the rotating frame.

We apply the Hill's transformation
\begin{equation} \label{EqHillTrans}
x=x_1+\mu^{1/3}\xi,\quad y=\mu^{1/3}\eta
\end{equation}
to equations (\ref{EqERequP}) and (\ref{EqERequB}) and, by letting $\mu\rightarrow 0$, we get (Ichtiaroglou, 1980)
\begin{equation} \label{EqEHequx1}
\ddot x_1 -x_{10}^4 x_1^{-3}+x_1^{-2}=0
\end{equation}
and
\begin{equation} \label{EqEHequ}
\begin{array}{l}
\ddot  \xi=2 x_{10}^2 x_1^{-2}\dot{\eta} + \left ( 2 x_1^{-3}+x_{10}^4 x_1^{-4}-\rho^{-3} \right ) \xi - 2 x_{10}^2 x_1^{-3}\dot{x}_1 \eta\\
\ddot \eta=-2 x_{10}^2 x_1^{-2}\dot{\xi} + \left (-  x_1^{-3}+x_{10}^4 x_1^{-4}-\rho^{-3} \right )\eta + 2 x_{10}^2 x_1^{-3}\dot{x}_1 \xi
\end{array}
\end{equation}
with $\rho^2=\xi^2+\eta^2$. 
The equation (\ref{EqEHequx1}) describes Keplerian motion with eccentricity, semi-major axis and period given by
\begin{equation}
e_p=x_{10}^3-1,\quad a=\frac{(1+e_p)^{1/3}}{(1-e_p)},\quad T=2 \pi \sqrt{\frac{(1+e_p)}{(1-e_p)^3}},
\end{equation}
respectively. For bounded motion, $x_{10}$ should be restricted in such a way that $-1<e_p<1$. For $e_p>0$ ($e_p<0$) the planet is at periapsis (apoapsis) at $t=0$ and it has $\dot{x}_1(0)=0$. For $e_p=0$ ($x_1=x_{10}=1$) we obtain the equations (\ref{EqCHillEqs}) of the CH model.

Fixing the value of the planetary eccentricity $e_p\neq 0$, the equations (\ref{EqEHequ}), which describe the motion of the massless satellite, constitute a periodic non-autonomous system of period $T$. The system (\ref{EqEHequ}) possesses only the symmetry $\mathbf{\Sigma}$ and an orbit of initial conditions
$$
\xi(0)=\xi_0, \quad \eta(0)=\dot{\xi}(0)=0, \quad \dot{\eta}(0)=\dot{\eta}_0
$$
is a {\em symmetric periodic orbit} of period $T'$ {\em if} it satisfies the periodicity conditions
$$
\dot{\xi}(T'/2;\xi_0,\dot{\eta}_0)=0,\quad \eta(T'/2;\xi_0,\dot{\eta}_0)=0.
$$
The period $T'$ has to be an integer multiple of the planetary period $T$, i.e. $T'=\kappa T$, $\kappa=1,2,..$.

By varying the planetary eccentricity $e_p$, a symmetric periodic orbit continues to exist, so that we get a monoparametric family (Ichtiaroglou, 1981). All periodic orbits that belong to such a family are isolated in phase space and have a period $T'$ which depends on $e_p$. For $e_p=0$, a family of the EH problem crosses a family of the CH problem. The crossing point must be a periodic orbit of the CH problem with period $T_c=T'/\lambda$, where $\lambda=1,2,..$ is the {\em multiplicity} of the periodic orbit. Since for $e_p=0$ it is $T=2\pi$, we conclude that the periodic orbits of the circular problem with period $T_c=2\kappa \pi/\lambda$ can be considered as ``bifurcation'' points for the EH problem. Therefore these orbits are continued for $e_p\neq 0$ with multiplicity $\lambda$ and period $T'$=$\lambda T_c$=$2\kappa \pi$. The ratio $\kappa/\lambda$ is the {\em resonance} of the periodic orbit.

Obviously, the number of bifurcation points on each family of the CH problem is infinite. In this study we restrict our attention to families that are of relatively small multiplicity and bifurcate mainly from stable periodic orbits. The linear stability of the orbits is determined by the two conjugate pairs of eigenvalues of the monodromy matrix of the system (\ref{EqEHequ}). An orbit is {\em stable} (s) only when all eigenvalues lie on the unit circle. If one or two pairs of eigenvalues lie on the real axis we have {\em single} (u) or {\em double} (uu) instability, respectively. Finally when all eigenvalues are outside the unit circle (but they are not real) we have {\em complex instability} (cu). The different kinds of stability can be determined by computing stability indices as in Broucke (1969) (see, also, Ichtiaroglou and Voyatzis, 1990). A necessary (but not sufficient) condition for a family of the EH problem to emanate (from $e_p=0$) with a stable branch is the bifurcation point to be a stable periodic orbit. Certainly, the stability type can change along the family.

\begin{figure}[htb]
\centering
$\begin{array}{ccc}
\includegraphics[width=5.5cm]{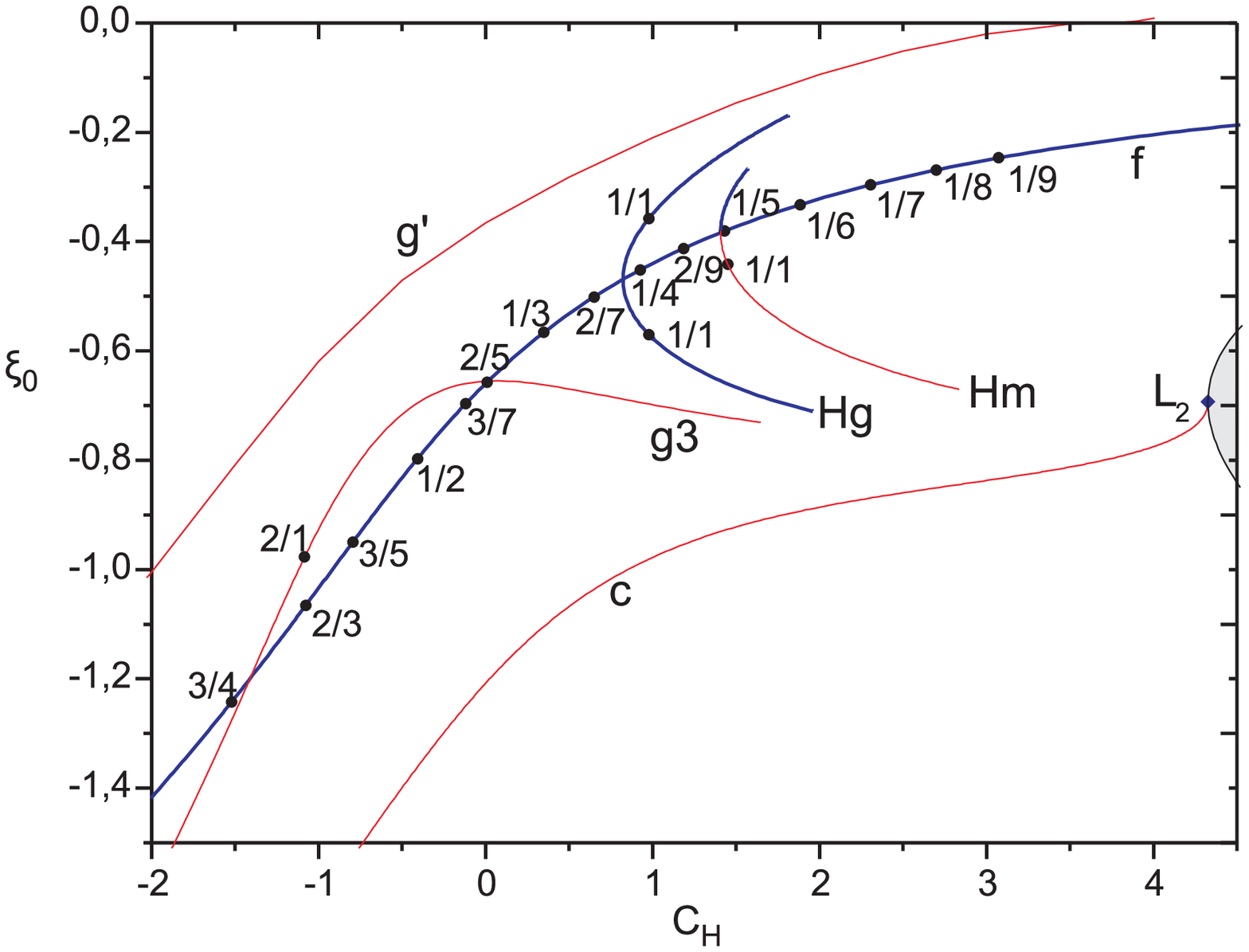} & \qquad &
\includegraphics[width=5.5cm]{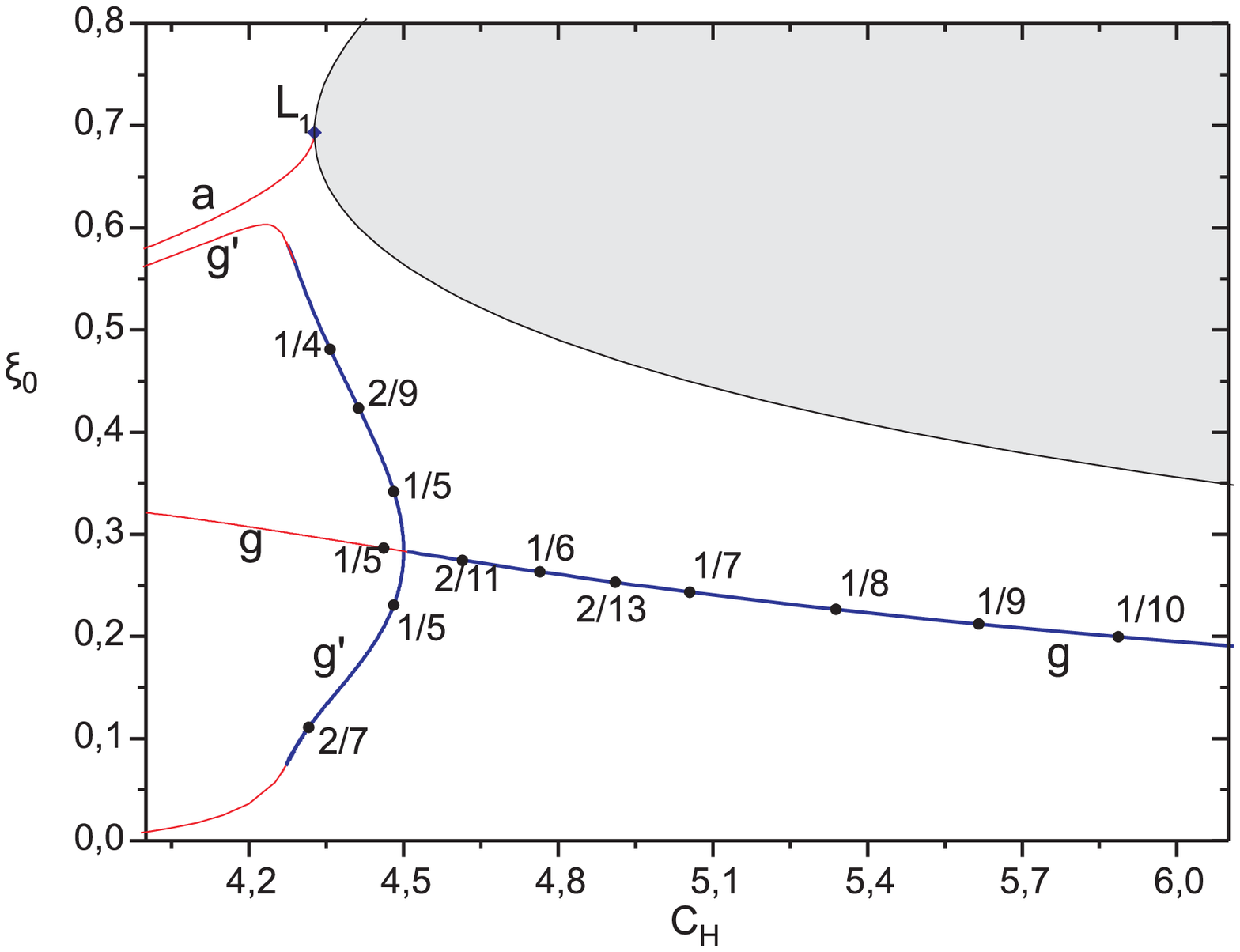}\\
\textnormal{(a)} & \qquad & \textnormal{(b)}
\end{array} $
\caption{Bifurcation points for continuation from the CH problem to the EH problem {\bf a} from retrograde families {\bf b} from prograde families of periodic orbits. The corresponding resonances are indicated.}
\label{FigBif}
\end{figure}

In Fig. \ref{FigBif} we present the bifurcation points that are studied in the present work. The resonance $\kappa/\lambda$ is indicated for each point. Most bifurcation points of Fig. \ref{FigBif} belong to the family $f$ of retrograde orbits (left panel) and to the stable parts of families $g$, $g'$ of prograde orbits (right panel). The bifurcation points of families $g3$, $Hg$ and $Hm$ play a special role, which will be discussed in the following section. We note that, as the Jacobi constant increases, the main families $f$ and $g$ tend to $\xi_0=0$ and the Hill approximation degenerates into the two-body problem.  In this case the bifurcation points to the EH model are of larger and larger multiplicity as $\xi_0\rightarrow 0$.

\section{Families of periodic orbits} \label{SecFamilies}
A monoparametric family of the EH problem, which emanates from a $\kappa/\lambda$ resonant bifurcation point, will be denoted as $F_e^{\kappa/\lambda}$, where $F$ indicates the name of the family of the CH model where the bifurcation point belongs to. For the circular problem we adopt the naming introduced by H\'{e}non. The families $a_e^{1/1}$, ${g'}_e^{1/1}$ and $f_e^{1/2}$ have been computed by Ichtiaroglou (1981) and are unstable (Ichtiaroglou and Voyatzis, 1990).

Each orbit of an EH family is represented by a point in the three dimensional space $\Pi_3=\{e_p,\xi_0,\dot{\eta}_0\}$ and a family forms a characteristic curve in this space. In the following we will present, for convenience, the families in the projection plane $(\xi_0,e_p)$. Thick (blue) or thin (red) characteristic curves indicate stable or unstable orbits, respectively. The periodic orbits have been computed by the method of differential corrections and satisfy the periodicity condition with an accuracy of $10^{-12}$. We always start from $e_p=0$, where the period and the initial conditions are known from the CH model.

\subsection{Families $f_e^{\kappa/\lambda}$ - retrograte orbits}

\begin{figure}[tb]
\centering
\includegraphics[width=9cm]{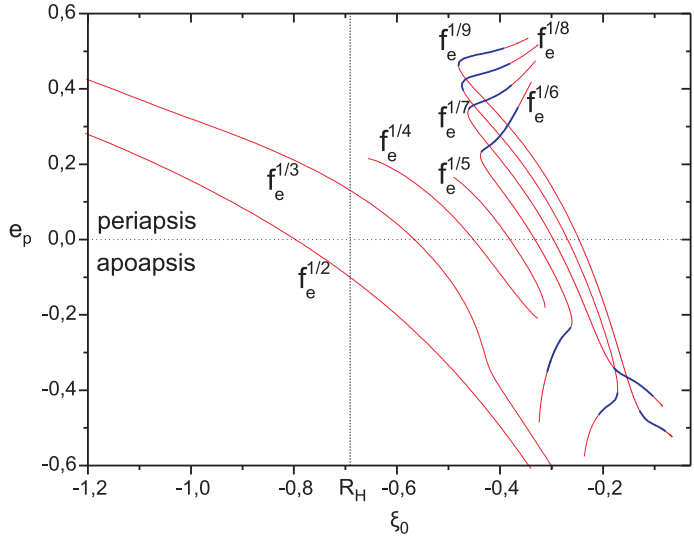}
\caption {Families $f_e^{1/\lambda}$ of periodic orbits of the EH problem. Thick (blue) or thin (red) curves indicate stable or unstable orbits, respectively.}
\label{Ffamf1}
\end{figure}

\subsubsection{The case $\kappa=1$}
We have computed the families up to multiplicity $\lambda=9$ and present them in Fig. \ref{Ffamf1}. The families $f_e^{1/2}$ and $f_e^{1/3}$ are unstable and extend outside of the Hill radius. The families with $6\leq \lambda \leq 9$ start (at $e_p=0$) as unstable but they posses a stable segment for moderate eccentricity values in both cases of the initial planetary position, namely for periapsis ($e_p>0$) and apoapsis ($e_p<0$).

\begin{figure}[htb]
\centering
$\begin{array}{ccc}
\includegraphics[width=5.5cm]{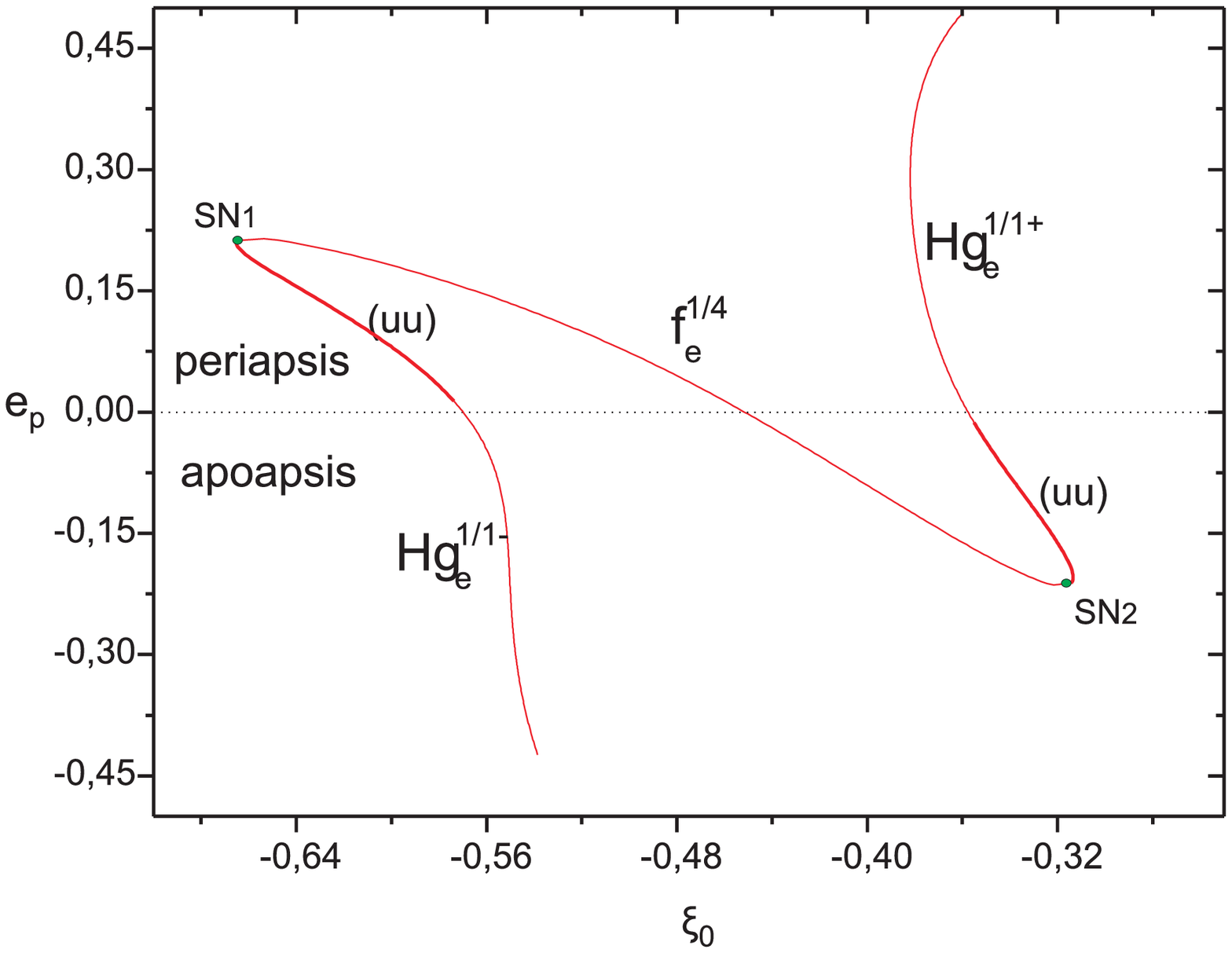} & \qquad &
\includegraphics[width=5.5cm]{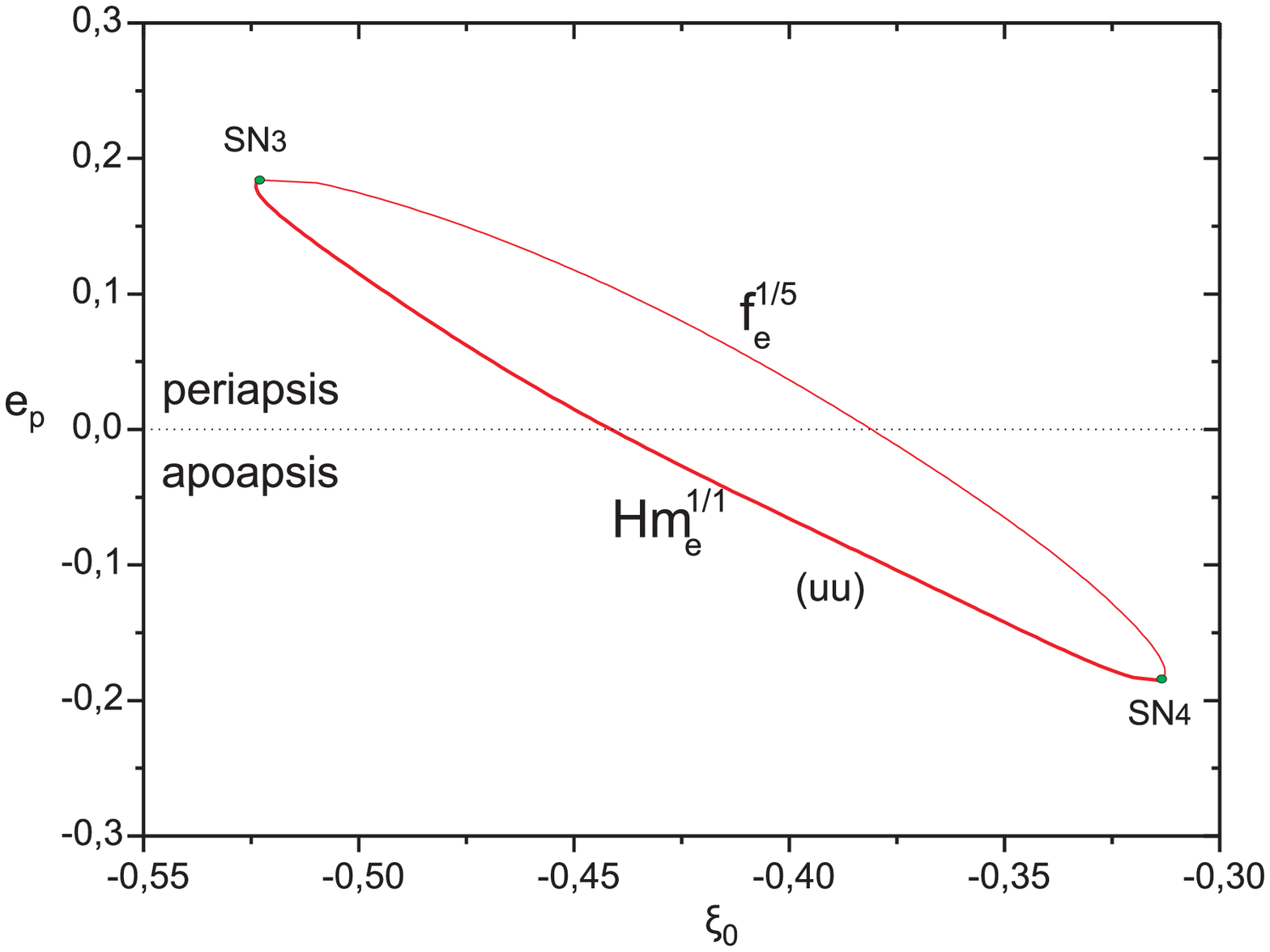}\\
\textnormal{(a)} & \qquad & \textnormal{(b)}
\end{array} $
\caption{ {\bf a} The continuation of family $f_e^{1/4}$ and its connection with families $Hg_e^{1/1}$.  {\bf b} The continuation of family $f_e^{1/5}$ and its connection with family $Hm_e^{1/1}$. A closed path is formed. The symbol (uu) indicates parts of the families with {\em double unstable} orbits.}
\label{Ffamf1plus}
\end{figure}

%%xxx revision
Families $f_e^{1/4}$ and $f_e^{1/5}$ in Fig. \ref{Ffamf1} seem to terminate abruptly at some critical points at $e_p=\pm 0.215$ and $e_p=\pm 0.185$, respectively. However, as it is shown in Fig. \ref{Ffamf1plus}, at these critical points, denoted as $SN$, saddle-node bifurcations take place. In these bifurcations the $1/1$ resonant families $Hg_e^{1/1\pm}$ and $Hm_e^{1/1}$, which bifurcate from the CH families $Hg$ of multiplicity $\lambda=4$ and $Hm$ of multiplicity $\lambda=5$, respectively, are involved. Particularly, in Fig. \ref{Ffamf1plus}a we obtain that the upper branch ($e_p>0$) of the family $f_e^{1/4}$ and the $Hg_e^{1/1-}$ emanate from the point $SN_1$, while the lower branch ($e_p<0$) of the family $f_e^{1/4}$ and the $Hg_e^{1/1+}$ emanate from $SN_2$. In Fig. \ref{Ffamf1plus}b we see that the families $f_e^{1/5}$ and $Hm_e^{1/1}$ emanate from both critical points $SN_3$ and $SN_4$, appearing for the eccentricity values $e_p=\pm 0.185$. The two families connect smoothly and form a closed characterisric curve.  

%%%%%%%%%%%%%%%%%%%%%%%%%%%

\

\begin{figure}[htb]
\centering
$\begin{array}{ccc}
\includegraphics[width=5.5cm]{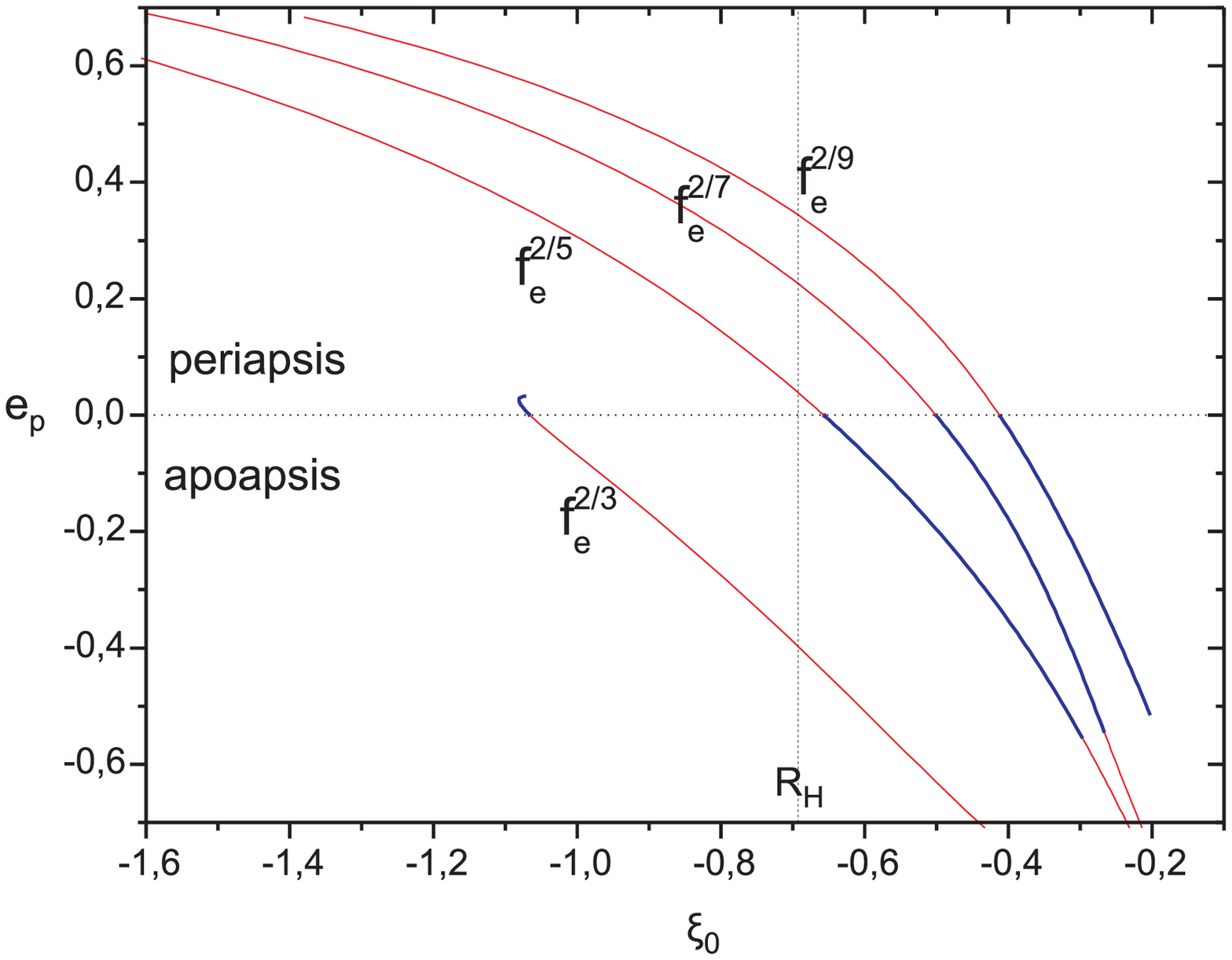} & \qquad &
\includegraphics[width=5.5cm]{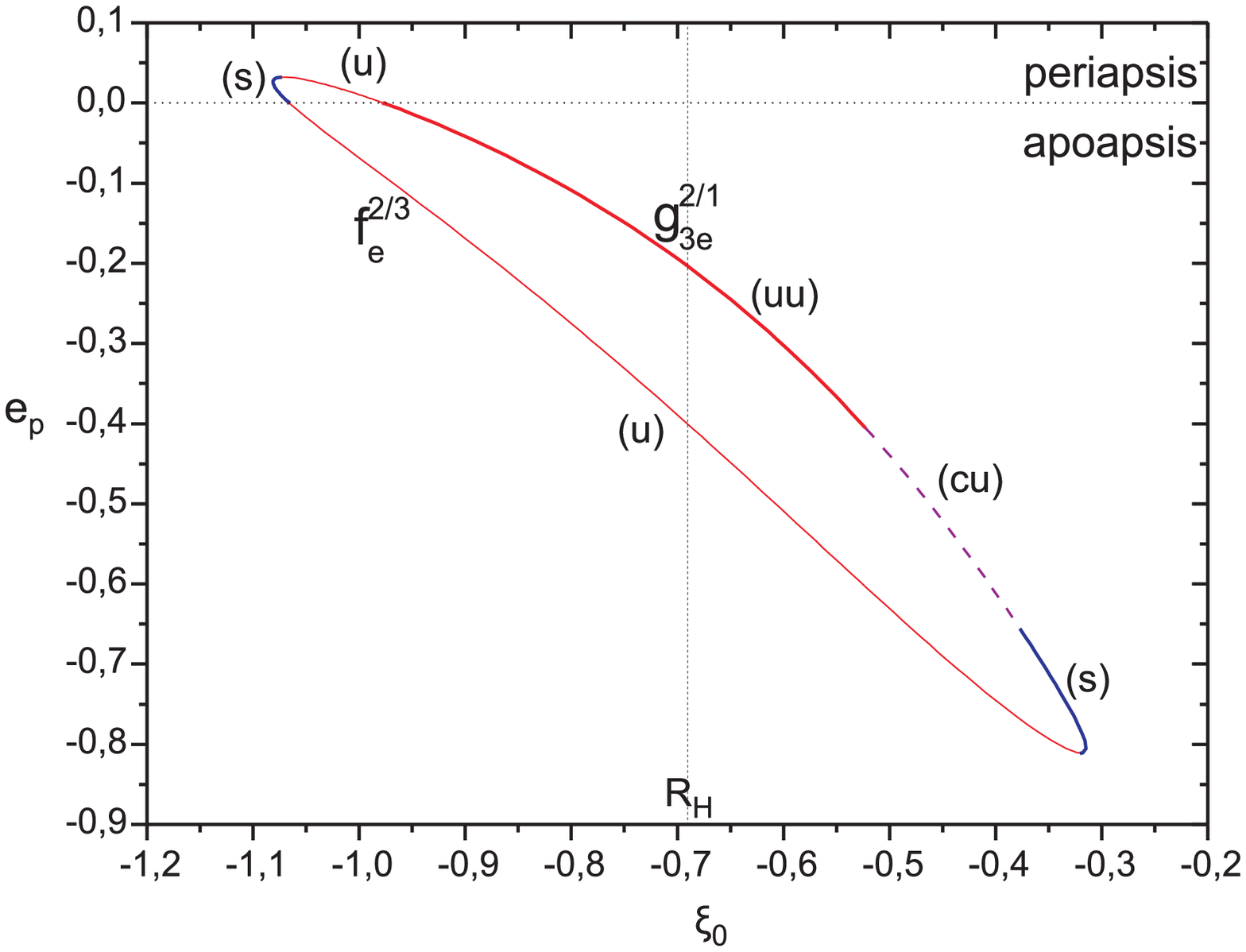}\\
\textnormal{(a)} & \qquad & \textnormal{(b)}
\end{array} $
\caption{ {\bf a} Families $f_e^{2/\lambda}$ of periodic orbits of the EH problem.  {\bf b} The continuation of family $f_e^{2/3}$ and its connection with family $g3_e^{2/1}$. Stability is indicated as (s) for stable (u) for unstable (uu) for double unstable and (cu) for complex unstable orbits.}
\label{Ffamf2}
\end{figure}

\subsubsection{The case $\kappa=2$}
The computed families are presented in Fig. \ref{Ffamf2}a. The families with $\lambda\geq 5$ are unstable for the periapsis case but they start as stable when the planet is initially at apoapsis. The stable segment extends up to high eccentricity values in all cases. In contrast, the family $f_e^{2/3}$ is unstable for $e_p<0$ and starts as stable for $e_p>0$. Then it seems to terminate but, as in the case of the family $f_e^{1/5}$, it connects with the family $g3_e^{2/1}$. The overall characteristic curve is given in Fig. \ref{Ffamf2}b. It forms a closed path, along which we find all possible types of linear stability. We note that the stable part is located outside the Hill sphere.

\begin{figure}[tb]
\centering
\includegraphics[width=7cm]{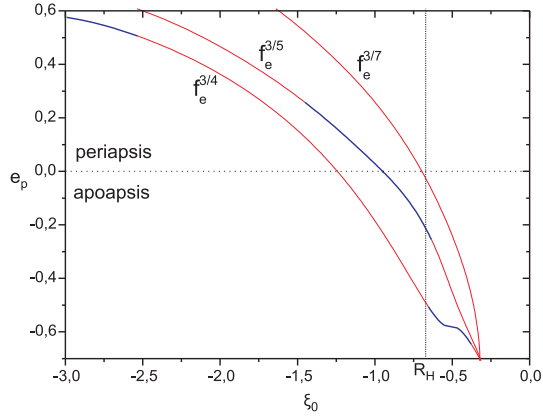}
\caption {Families $f_e^{3/\lambda}$ of periodic orbits of the EH problem. Thick (blue) or thin (red) curves indicate stable or unstable orbits, respectively.}
\label{Ffamf3}
\end{figure}

\subsubsection{The case $\kappa=3$}
We computed the families  $f_e^{3/4}$, $f_e^{3/5}$ and $f_e^{3/7}$ presented in Fig. \ref{Ffamf3}. All the corresponding bifurcation points are located outside the Hill sphere. The family $f_e^{3/4}$ is unstable for small and moderate values of the planetary eccentricity, but it posses stable segments for $|e_p|>0.5$. We note that for $e_p<0$ the family becomes stable when it enters the Hill sphere. On the other hand, the family $f_e^{3/5}$ is stable approximately in the interval $-0.25<e_p<0.25$. Now, the family becomes unstable when it enters the Hill sphere for $e_p<0$. The family $f_e^{3/7}$  is unstable in all the explored domain.

\subsection{Families $g_e^{\kappa/\lambda}$ and ${g'}_e^{\kappa/\lambda}$ - prograte orbits}
In this case all bifurcation points and families are located inside the Hill sphere.

\begin{figure}[htb]
\centering
$\begin{array}{ccc}
\includegraphics[width=5.5cm]{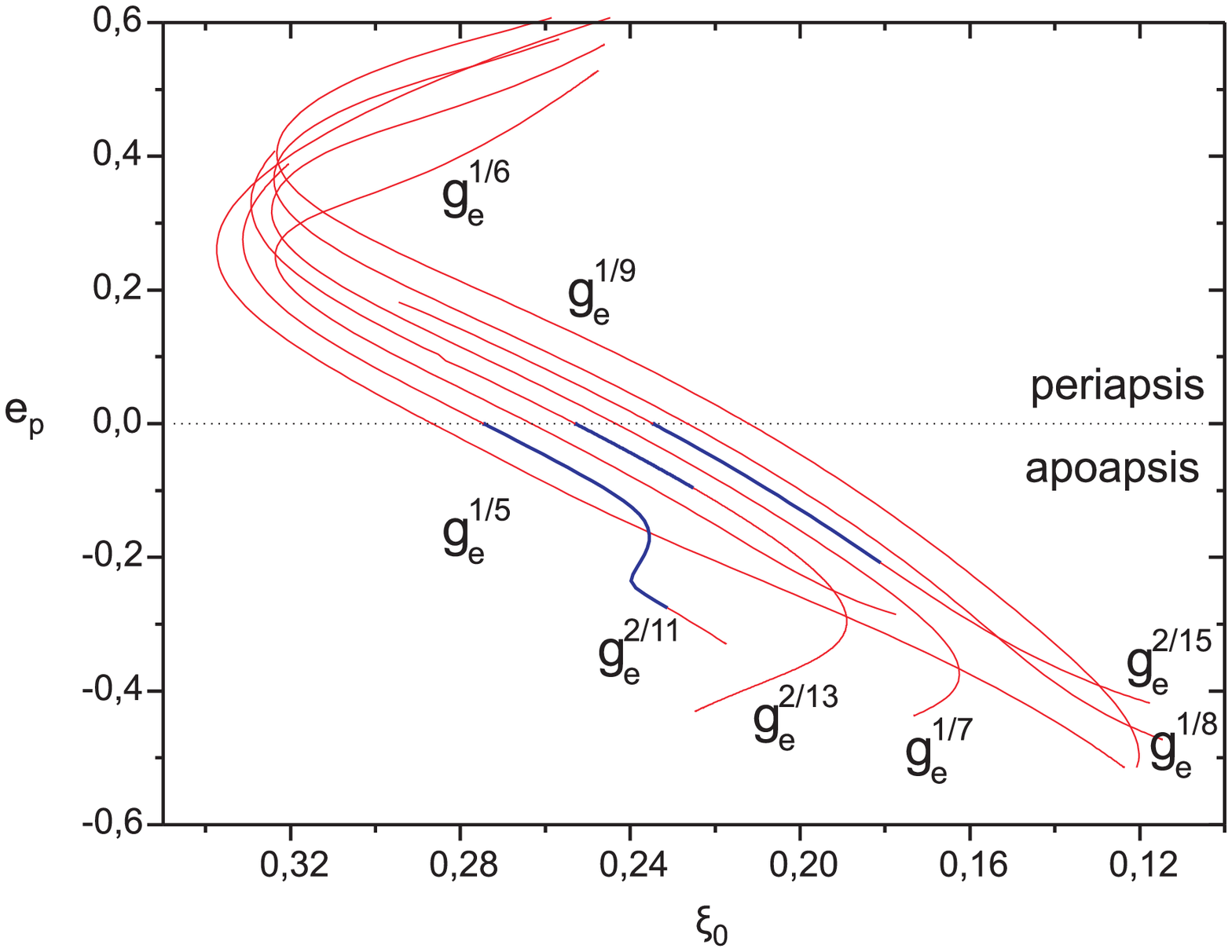} & \qquad &
\includegraphics[width=5.5cm]{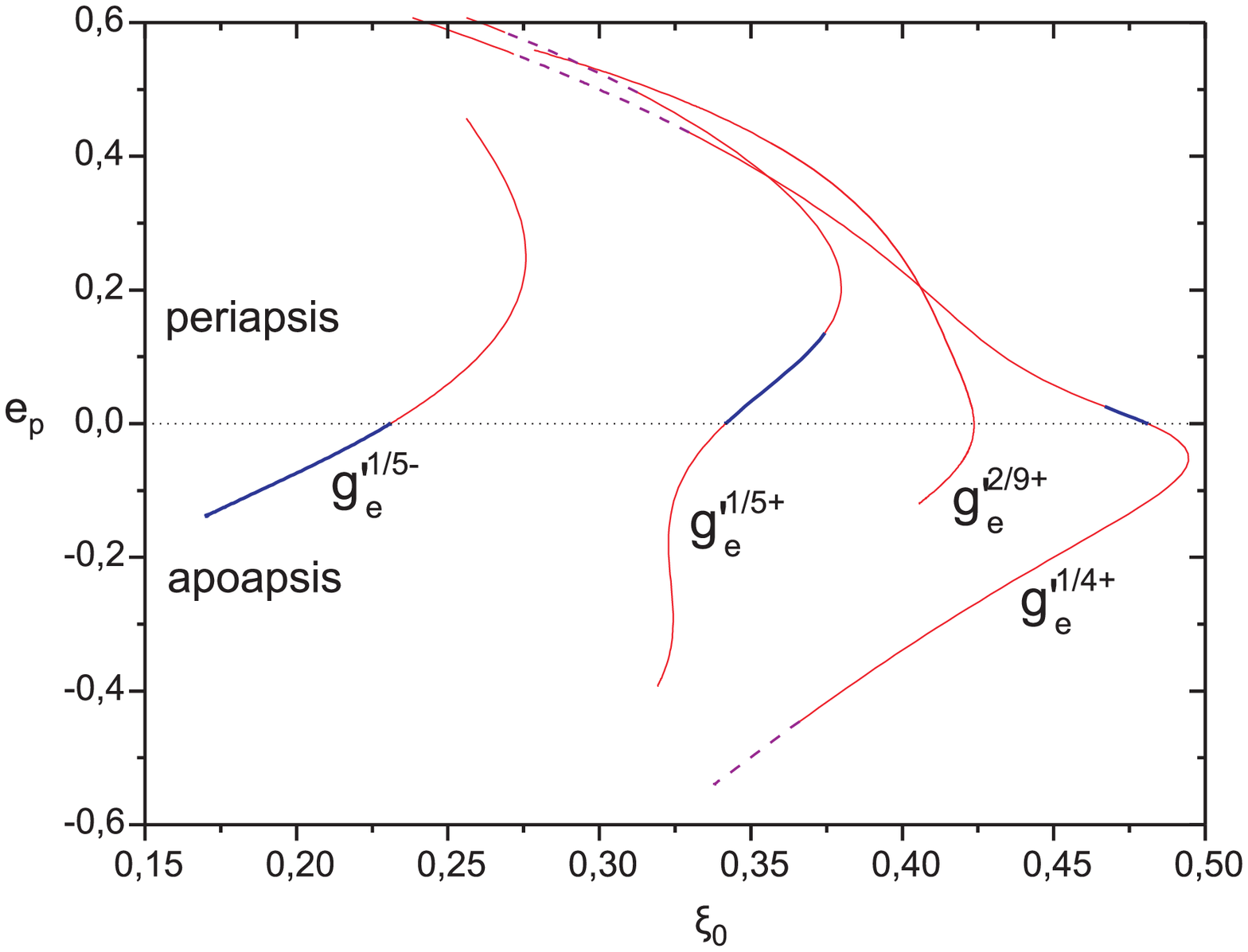}\\
\textnormal{(a)} & \qquad & \textnormal{(b)}
\end{array} $
\caption{ {\bf a} Families $g_e^{\kappa/\lambda}$   {\bf b}  Families ${g'}_e^{\kappa/\lambda}$. Thick (blue) or thin (red) solid curves indicate stable or simply unstable orbits, respectively, and dashed curves correspond to complex unstable orbits. }
\label{FfamG}
\end{figure}

\subsubsection{Families $g_e^{\kappa/\lambda}$}
We computed the families $g_e^{1/\lambda}$ with $5\leq\lambda\leq 9$ and the families  $g_e^{2/\lambda}$ with $\lambda=11$, 13 and 15. They are presented in Fig. \ref{FfamG}a. For $e_p>0$ all families are unstable. For $e_p<0$ the families $g_e^{1/\lambda}$ are also unstable but the families $g_e^{2/\lambda}$ start us stable and become unstable when the planetary eccentricity exceeds a critical value, which is different for each case.

\subsubsection{Families ${g'}_e^{\kappa/\lambda}$}
The CH family $g'$ has two branches (see Fig. \ref{FigCHfams}) and each resonant bifurcation point appears twice, one in each branch; we use the notation ``$+$'' and ``$-$'' to distinguish between them. In this case the convergence of our numerical approach becomes delicate. The families we were able to compute are presented in Fig. \ref{FfamG}b. Also in this case we see that segments of stable orbits exist for low eccentricities.

\section{Phase space numerical exploration}
\begin{figure}[tb]
\centering
\includegraphics[width=7cm]{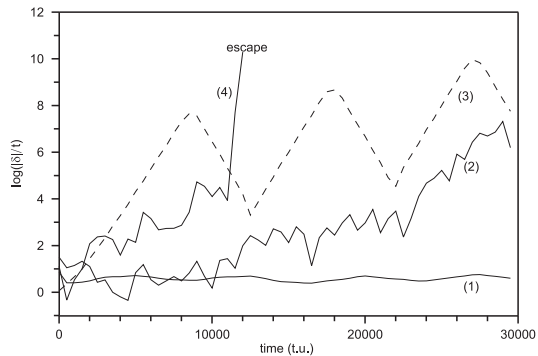}
\caption {The evolution of the FLI (without including the supremum) for some typical cases. Case (1) corresponds to a regular orbit, case (2) to a chaotic orbit and case (4) to a chaotic orbit that escapes for $t<t_{max}$. Case (3) refers to an unstable periodic orbit (see the text).}
\label{FigFLIevol}
\end{figure}

In this section we study the evolution of generally non-periodic orbits, associated with particular grids of initial conditions and various values of the planetary eccentricity, $e_p$. The main orbit classification aims at distinguishing the orbits between bounded and escaping ones. We set the escape criterion to $\rho>15\,R_H$. Additionally, along the numerical integration of the orbit, we compute a Fast Lyapunov Indicator (FLI, Froeschl\'e and Lega, 2000), in order to classify bounded orbits as regular or chaotic. The FLI, as defined by Voyatzis (2008), is given by the relation
\begin{equation}
FLI=\sup_{t\leq t_{max}} \frac{1}{t}|\delta(t)|,
\label{EqFLIndic}
\end{equation}
where $\delta(t)$ is the deviation vector, whose evolution is given by the solution of the variational equations. In Fig. \ref{FigFLIevol} we present the behaviour of the FLI for some typical trajectories. Case (1) corresponds to a regular (quasi-periodic) orbit, while case (4) corresponds to an irregular orbit that escapes at some $t<t_{max}$. Cases (2) and (3) are rather rare. Case (2) corresponds to a  chaotic orbit which is sticky and does not escape for $t<t_{max}$. Case (3) corresponds to a trajectory that lies very close to an unstable periodic orbit. It is a typical case for all unstable periodic orbits presented in the previous section with relatively small planetary eccentricity ($e_p\lessapprox 0.2$). When the orbit follows the stable (unstable) manifold of the periodic orbit, the FLI decreases (increases) (see also Skokos et al., 2007). The overall (average) FLI evolution shows a linear increase in logarithmic scale, which is an indication of the existing instability.

We have computed stability maps corresponding to a $N\times M$ grid of initial conditions. For each orbit numerical integration is performed up to $t_{max}=30000$ time units or $FLI<FLI_{max}$(=$10^{10}$) or until the escape condition is fulfilled. Strongly chaotic orbits escape in relatively short time intervals. We observed that even if an orbit reaches the maximum FLI value, it shows a weakly chaotic evolution, at least for $t<t_{max}$.       

\begin{figure}[tb]
\centering
\includegraphics[width=9cm]{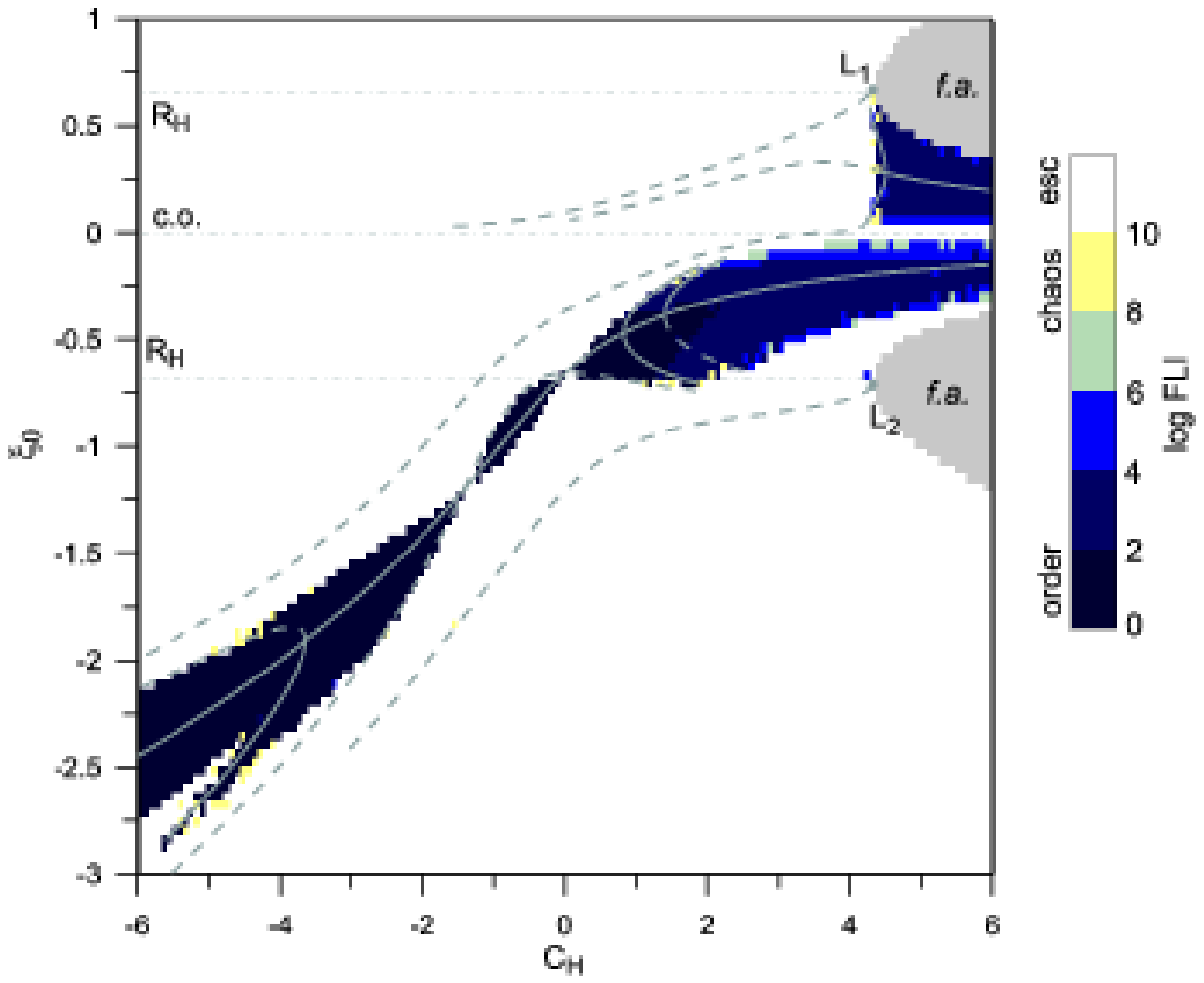}
\caption {H\'enon stability map for the CH model ($e_p=0$). The orbits of the grid are classified according to the color bar on the right of the map.  The main families of periodic orbits are also presented with solid (stable) and dashed (unstable) curves. The two horizontal lines $R_H$ determine the boundaries of the Hill's sphere. For $\xi_0=0$ we get collision orbits (c.o.) while the grey regions (f.a.) indicate forbitten regions.}
\label{FigGridHenon0}
\end{figure}

\begin{figure}
\centering
$\begin{array}{ccc}
\includegraphics[width=5.5cm]{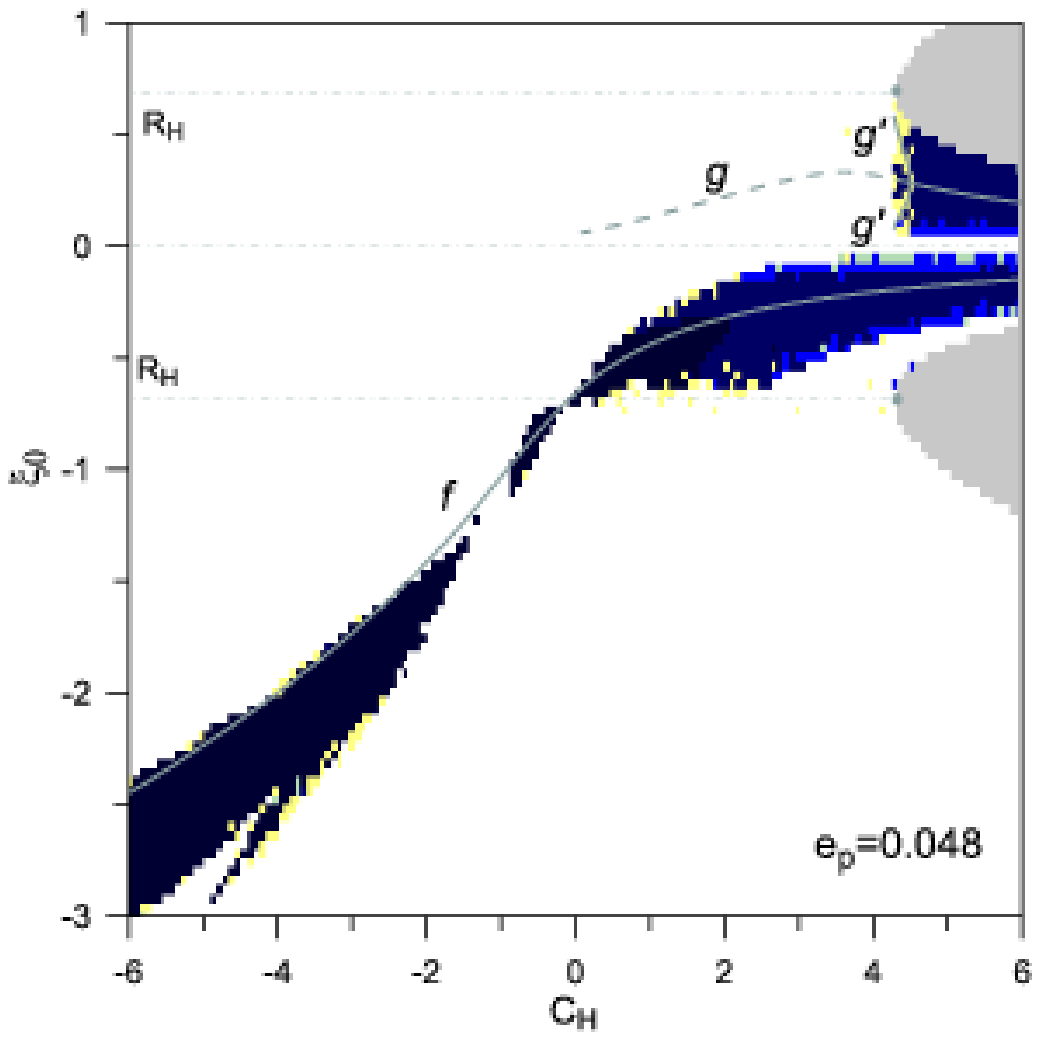} & \qquad & \includegraphics[width=5.5cm]{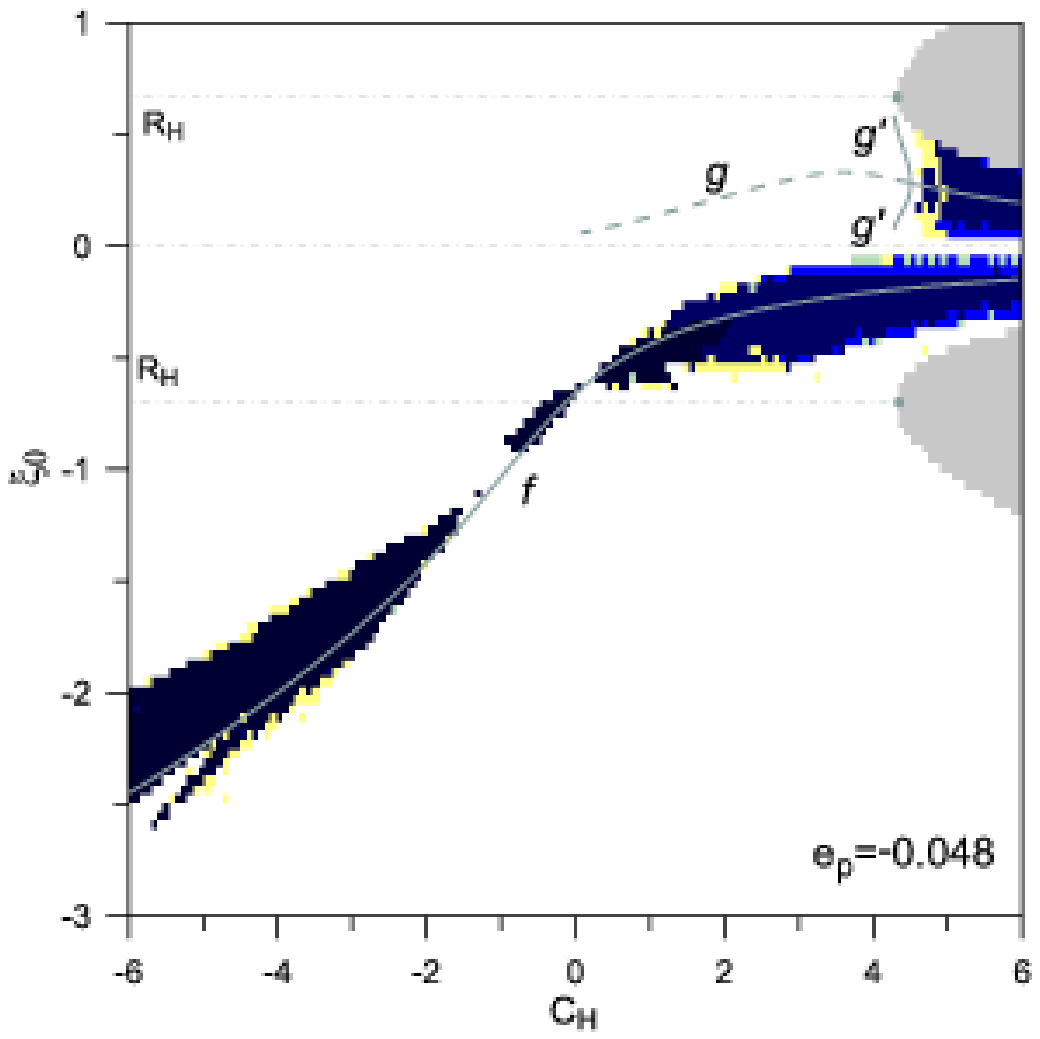}\\
\textnormal{(a)} & \qquad & \textnormal{(b)} \\
\includegraphics[width=5.5cm]{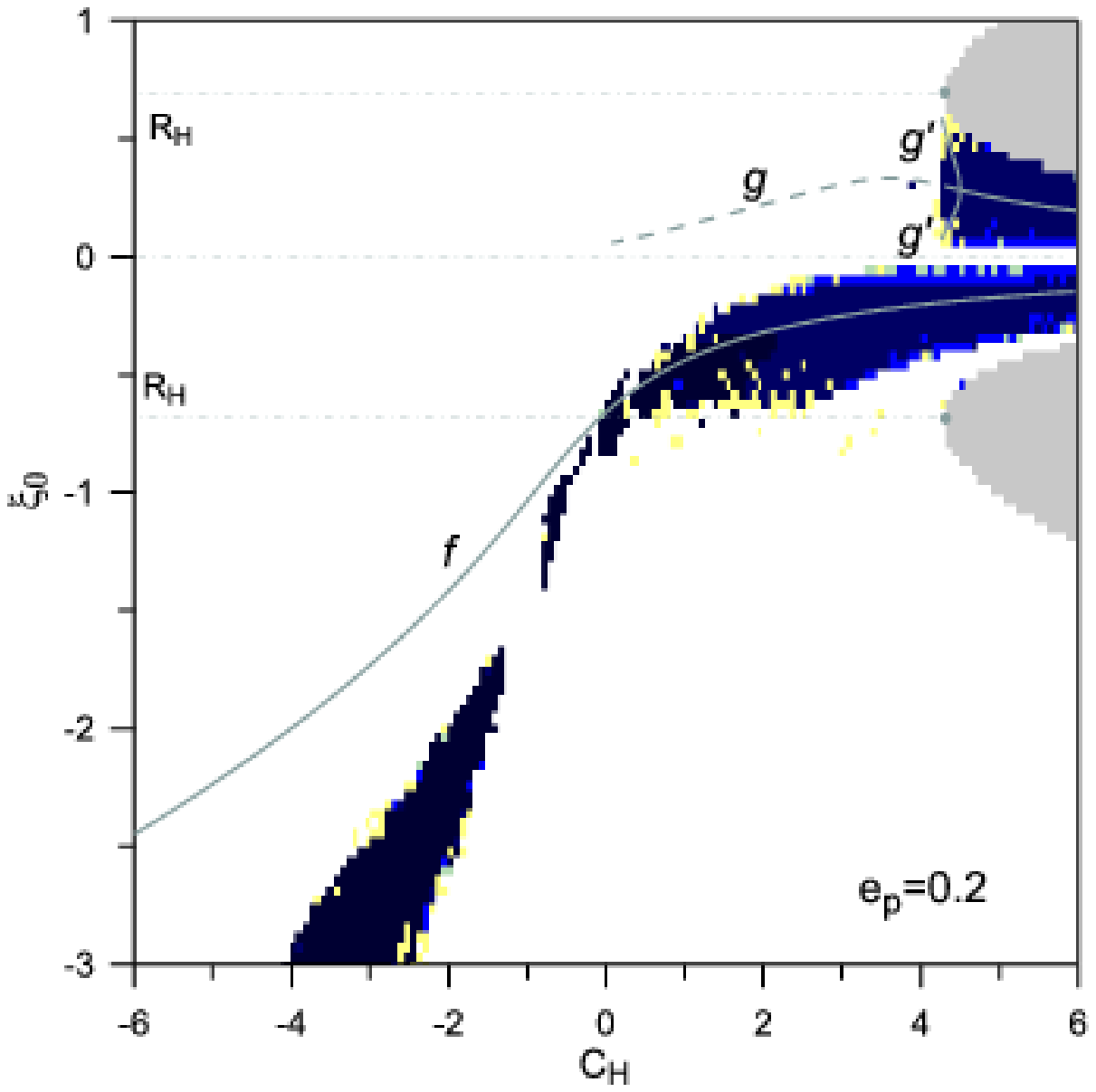} & \qquad & \includegraphics[width=5.5cm]{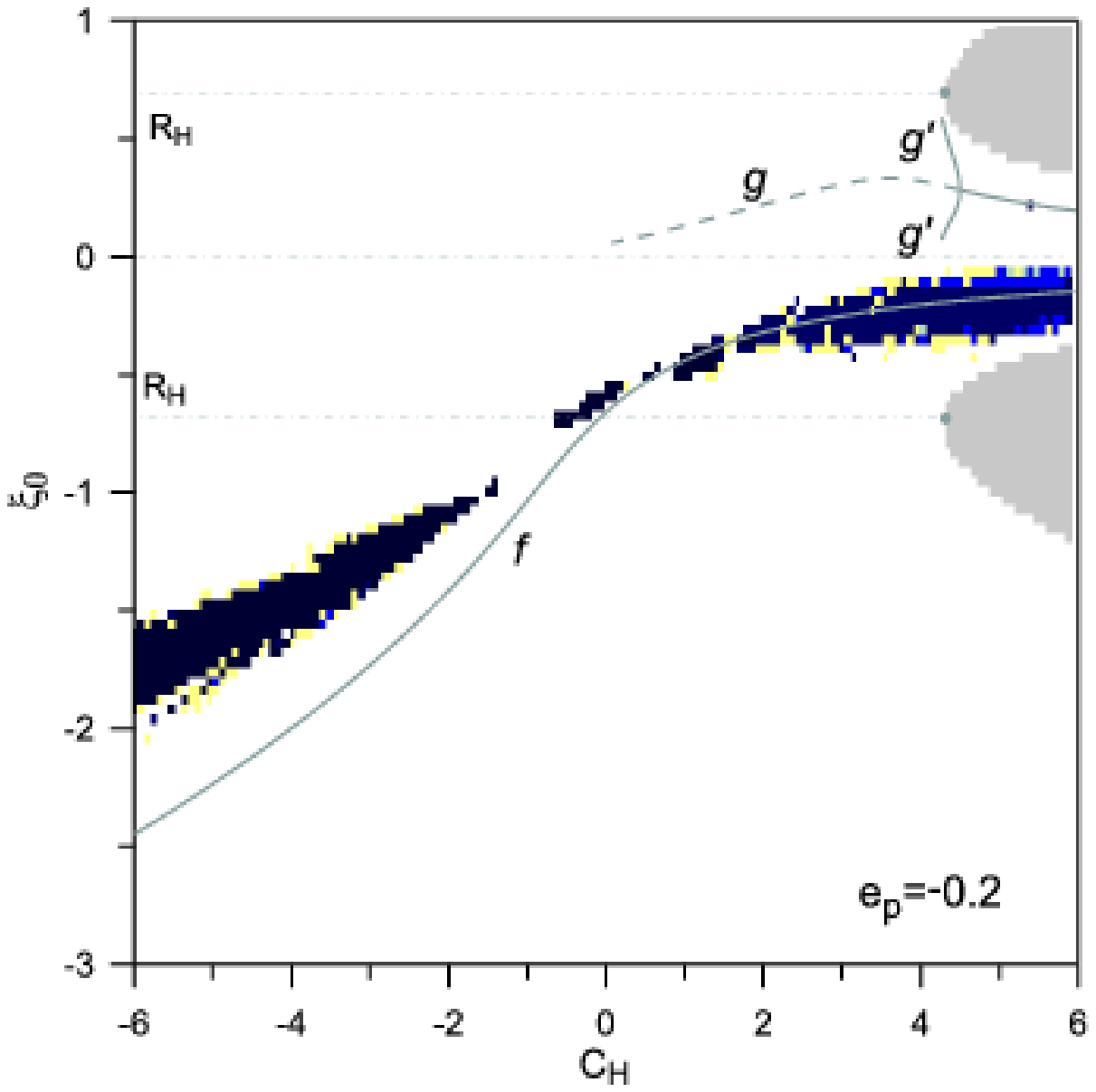}\\
\textnormal{(c)} & \qquad & \textnormal{(d)} \\
\includegraphics[width=5.5cm]{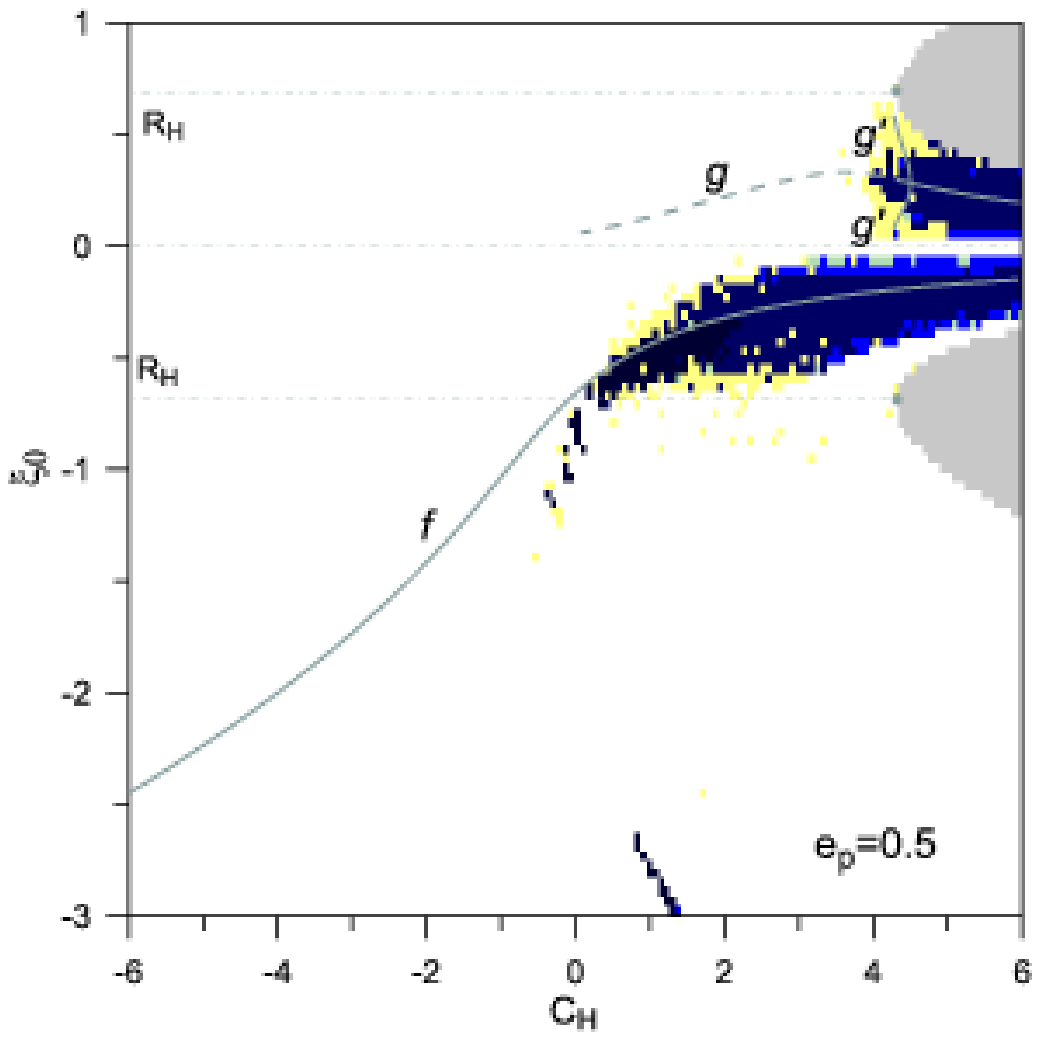} & \qquad & \includegraphics[width=5.5cm]{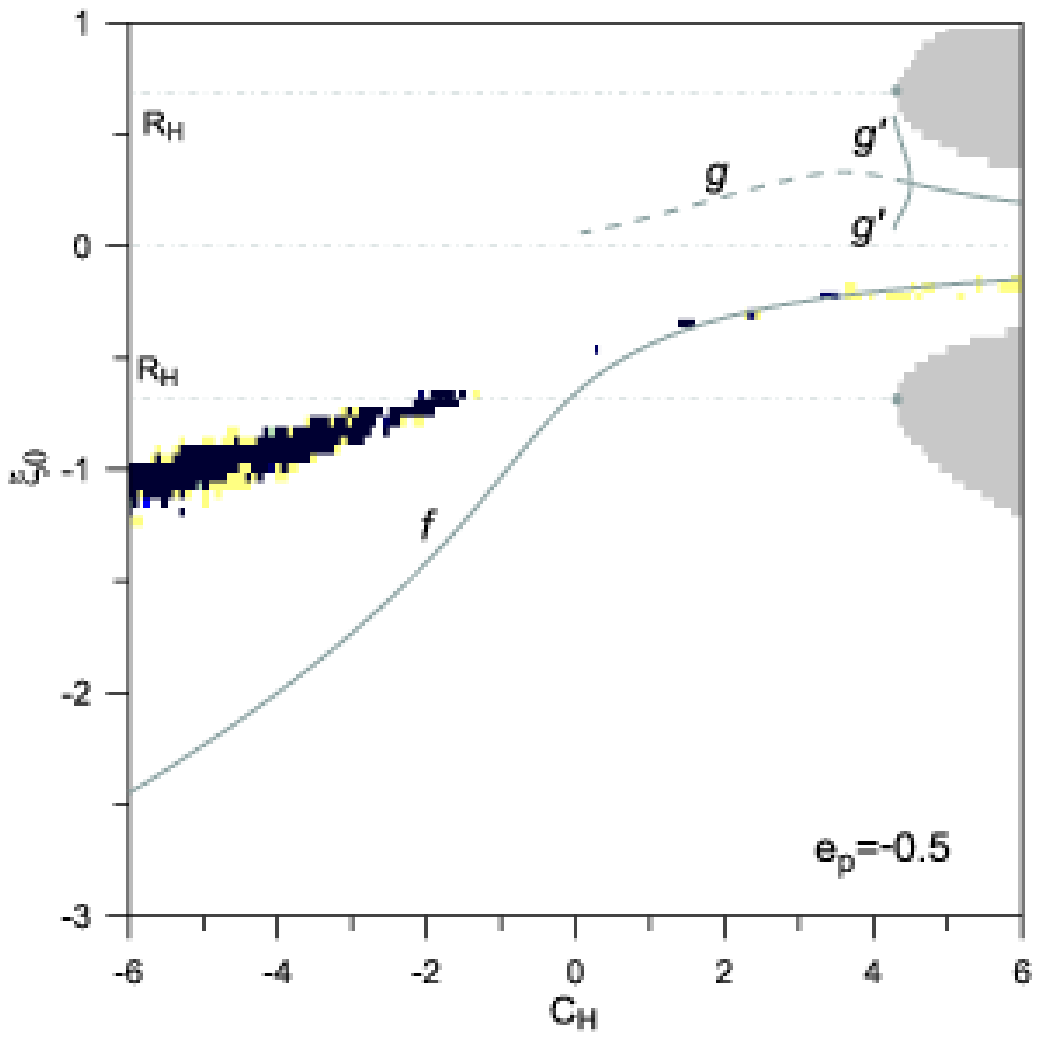}\\
\textnormal{(e)} & \qquad & \textnormal{(f)}
\end{array} $
\caption{H\'enon stability maps of grid size $150\times 100$ for the indicated planetary eccentricity values. Panels on the left correspond to the {\em periapsis} case and panels on the right to the {\em apoapsis} case. The color map is as in Fig. \ref{FigGridHenon0}.}
\label{FigGridHenonEH}
\end{figure}

\subsection{H\'enon stability maps}
In the circular Hill model, which has two degrees of freedom, the method of Poincar\'e section can depict clearly the topology of phase space and the qualitative characteristics of the trajectories (H\'enon, 1970; Chauvineau and Mignard, 1991). For $C_H>3^{4/3}$ the zero velocity curves define closed regions on the $O\xi\eta$ plane and all orbits, either regular or chaotic, are bounded. For $C_H<3^{4/3}$ the majority of chaotic orbits are escape orbits. H\'enon (1970) determined the borders of bounded motion by considering the plane $C_H-\xi_0$ of initial conditions, with $\eta_0=0$, and $\dot \eta_0>0$, which are defined by the Jacobi integral (\ref{EqJacobi}).  This diagram of bounded motion, which is called {\em H\'enon diagram}, has been used also by Shen and Tremaine (2008) in order to obtain bounded satellite motion by using numerical integrations of a 6-body model, which includes the Sun, the four giant planets and the satellite as a massless body. The H\'enon diagrams showed that regions of bounded motion are located near the main families ($f$, $g$ and $g'$) of stable periodic orbits.

Following our methodology, the H\'enon map (or diagram) for the CH model ($e_p=0$) is given in Fig. \ref{FigGridHenon0}. Empty (white) regions correspond to escape orbits. The regular orbits (dark regions) are distributed around the stable families of periodic orbits and this is in agreement with the results of H\'enon (1970). The reader can observe the tangle of stability around the stable family $Hn$, and this fact indicates the efficiency of the FLI method to provide detailed stability maps. Chaotic orbits (say orbits with $FLI\gtrapprox10^6$) are almost absent in the map, since the majority of chaotic orbits escape during the numerical integration.

Using the same initial conditions as those of Fig. \ref{FigGridHenon0}, we set $e_p\neq 0$ and repeat the computations. Note that positive or negative values of $e_p$ indicate that the planet is initially at periapsis ($p$-case) or  apoapsis ($a$-case), respectively.  Also, we remark that the points of the characteristic curves $f$, $g$ and $g'$ do not correspond to periodic orbits in this case. 
In Figs. \ref{FigGridHenonEH}a and \ref{FigGridHenonEH}b we present the stability maps for the $p$ and $a$ case, respectively, and for $e_p=0.048$ (Jupiter's eccentricity). We can observe that the width of the stable region of retrograde motion (i.e. around the family $f$) has shrunk but not significantly. The tail of this region ($C_H<0$) seems now to be located below the characteristic curve $f$ for the $p$-case and above it for the $a$-case. A rather significant effect of the non-zero planetary eccentricity value is observed in case of prograde orbits and mainly those located close to the $g'$ family, where chaotic orbits appear. This is more clear in the $a$-case, where now the ordered orbits are obtained for higher values of $C_H$, compared to those of the circular case.

By increasing the planetary eccentricity to $e_p=0.2$ (periapsis case), we see from the corresponding map of Fig. \ref{FigGridHenonEH}c that the area of the stability regions remains practically unchanged for retrograde as well as prograde orbits. However, for the apoapsis case ($e_p=-0.2$) the stability region of retrograde orbits has shrunk significantly, while the stability region of prograde orbits has disappeared completely from the map (actually stability can be found for higher values of $C_H$). Finally in Figs. \ref{FigGridHenonEH}e and \ref{FigGridHenonEH}f the case of a relatively high eccentricity value $(|e_p|=0.5)$ is shown. It is remarkable that in the $p$-case we can still find regions of stable orbits for both retrograde and prograde cases. But retrograde stable orbits are found almost only inside the Hill's sphere. In contrast, the $a$-case shows an extensive reduction of the area of stable orbits. The only significant region of stable (retrograde) orbits left is located outside the Hill sphere.

\subsection{Stability maps along CH families}
The H\'enon maps reveal that, in the case of the EH model, regions of stable satellite orbits are located around the stable families of the CH model. The effect of the planetary eccentricity to the periodic orbits of a given family, say $A$, of the circular model can be studied by constructing stability maps on the plane $\xi_0 - e_p$, where for each value $\xi_0$ we complete the set of initial conditions with $\eta_0=\dot\xi_0=0$ and $\dot\eta_0$ has the value that corresponds to the particular periodic orbit of the given family $A$. Thus, the initial conditions of the line $e_p=0$ correspond exactly to the family $A$ of periodic orbits.

\begin{figure}[tb]
\centering
\includegraphics[width=8cm]{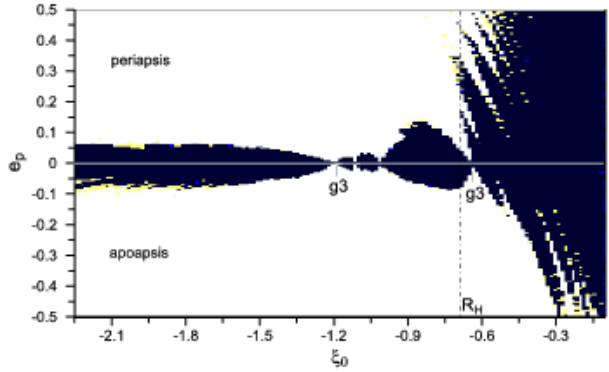}
\caption {Stability map ($150\times 200$ grid) along the stable family $f$ of retrograde orbits. The color map is as in Fig. \ref{FigGridHenon0}.}
\label{FigGridCHfamf}
\end{figure}

The stability map along the family $f$ is shown in Fig. \ref{FigGridCHfamf}. Inside the Hill's sphere ($-3^{-1/3}<\xi_0<0$) the stable orbits dominate and exist up to relatively high eccentricity values. Outside the Hill's sphere, we obtain a region of stability around the family $f$ ($e_p=0$). Two of the gaps in this region are associated with the crossing of the unstable family $g3$ by the family $f$ and can be also seen in the H\'enon maps of Figs. \ref{FigGridHenon0} and \ref{FigGridHenonEH}a,b. This region is restricted to low (absolute) eccentricity values, because we have seen that, as the eccentricity increases, the stable regions located outside the Hill's sphere move away from the characteristic curve $f$.

\begin{figure}[tb]
\centering
$\begin{array}{ccc}
\includegraphics[width=5.5cm]{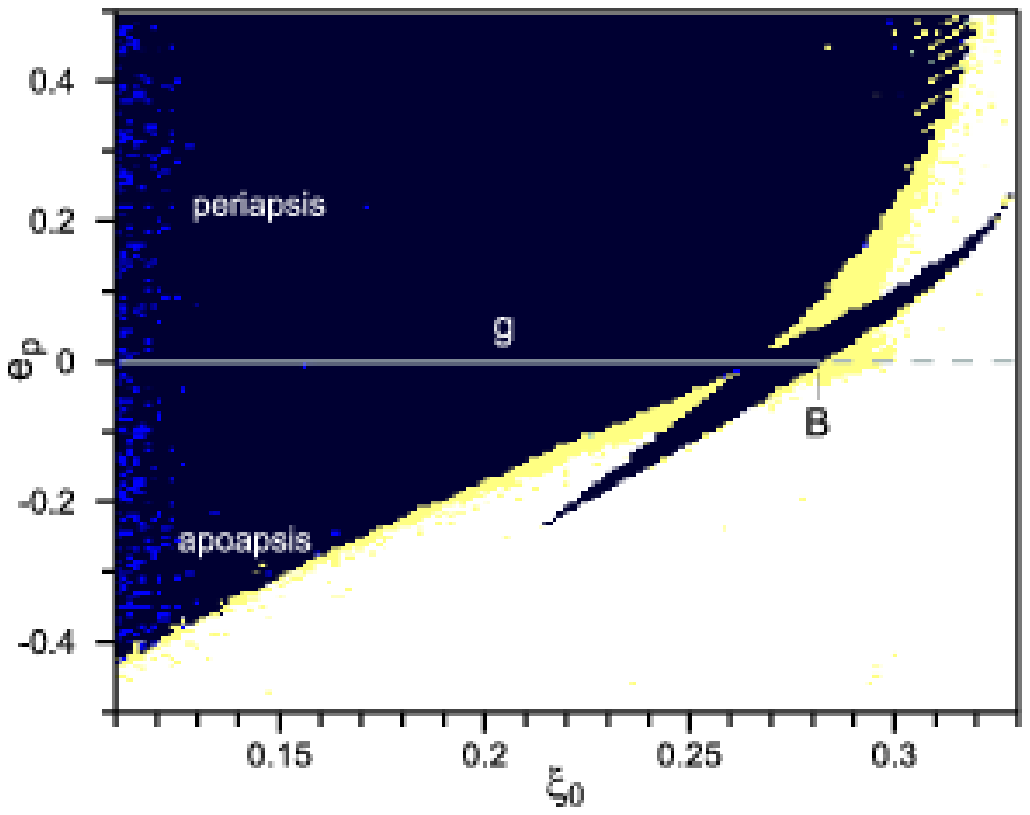} & \qquad & \includegraphics[width=5.5cm]{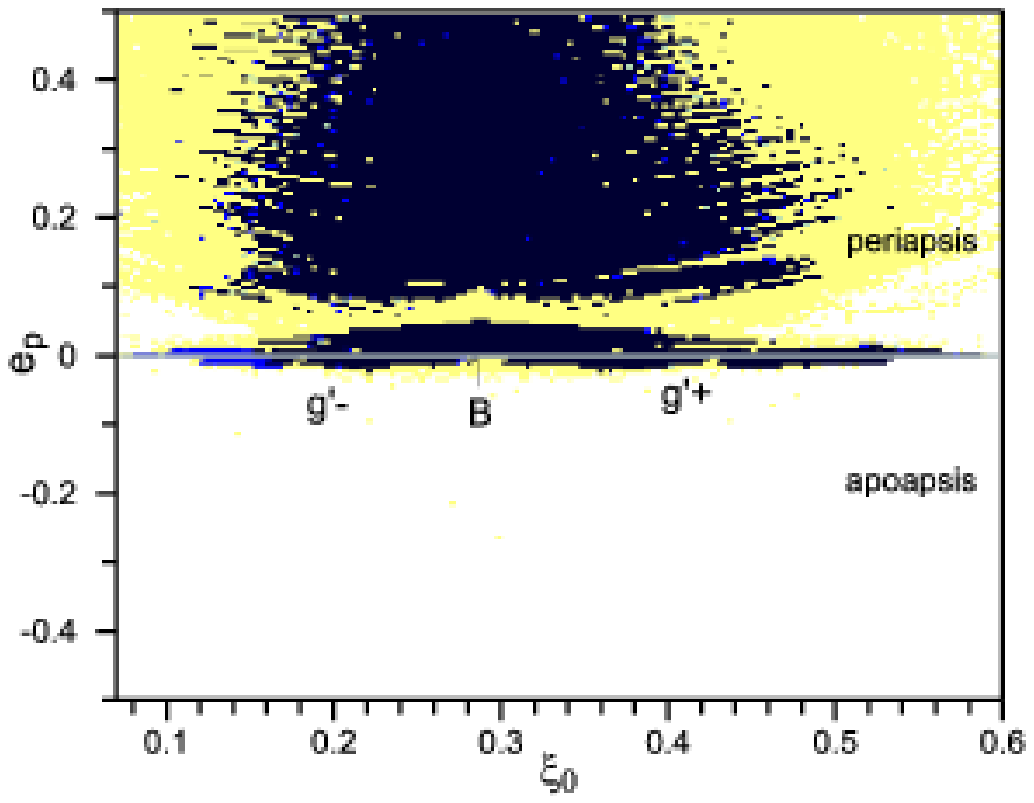}\\
\textnormal{(a)} & \qquad & \textnormal{(b)} \\
\end{array} $
\caption{Stability maps ($150\times 200$ grid) along {\bf a.} family $g$ {\bf b.} family $g'$. The color map is as in Fig. \ref{FigGridHenon0}.}
\label{FigGridCHfamg}
\end{figure}

The stability map along the family $g$ is shown in Fig. \ref{FigGridCHfamg}a. The point B indicates the pitchfork bifurcation of the family $g'$ from the family $g$ (H\'enon, 1969). No stable orbits are found around the unstable part of the family (on the right of the bifurcation point B). Around the stable part of the family, regions of ordered motion dominate, but the asymmetry between the periapsis and apoapsis cases is clearly seen. A similar asymmetry is also seen in the map along the family $g'$, shown in Fig. \ref{FigGridCHfamg}b. So we conclude that the {\em periapsis} case gives significantly more stable prograde orbits, compared to the {\em apoapsis} case, in agreement with the results obtained from the H\'enon maps.

\subsection{Stability maps along EH families}
In the previous cases we associated regions of stable motion of the EH model with the families of periodic orbits of the CH model. Additionally, new regions of stability are expected to exist around stable periodic orbits of the EH model. We construct stability maps centered along a given family $F$ of periodic orbits and defined by grids of initial conditions on the plane $e_p$-$\Delta\xi_0$. Each point of such a grid corresponds to the initial conditions $\bar{\xi}_0+\Delta\xi_0$, $\eta_0=\dot\xi_0=0$ and $\bar{\dot\eta}_0$, where ($\bar{\xi}_0$, $\bar{\dot\eta}_0$) are the initial conditions of the periodic orbit of family $F$ at the planetary eccentricity $e_p$, which has a period $T'$. Thus, the family $F$ is located on the axis $\Delta\xi_0=0$. For these maps we have set $t_{max}=20000 T'$ and $FLI_{max}=10^{20}$.

\begin{figure}[tb]
\centering
$\begin{array}{ccc}
\includegraphics[width=5.5cm]{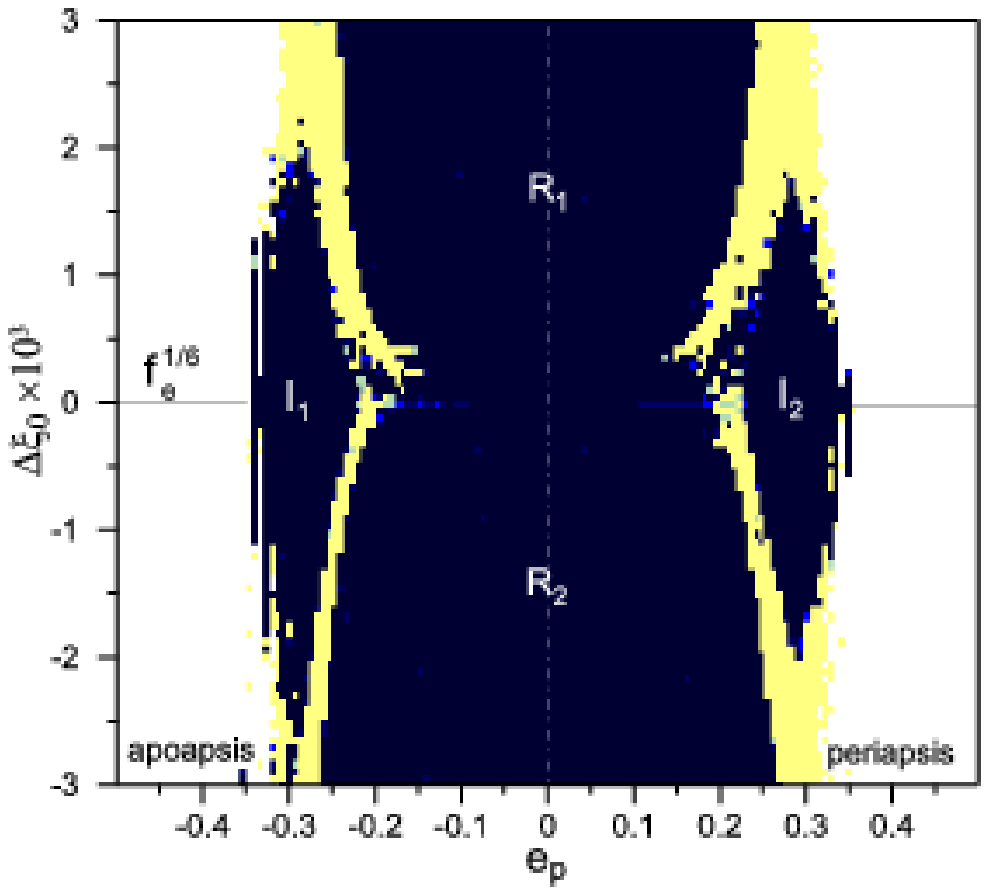} & \qquad & \includegraphics[width=5.5cm]{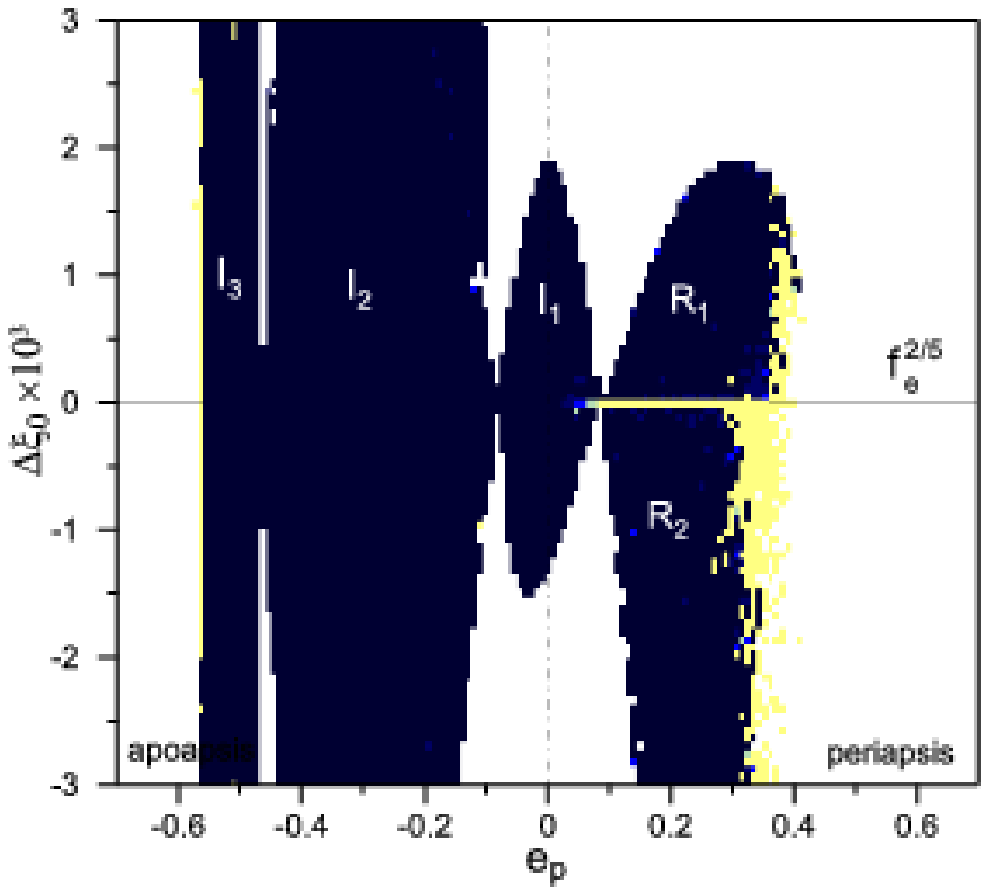}\\
\textnormal{(a)} & \qquad & \textnormal{(b)} \\
\includegraphics[width=4cm]{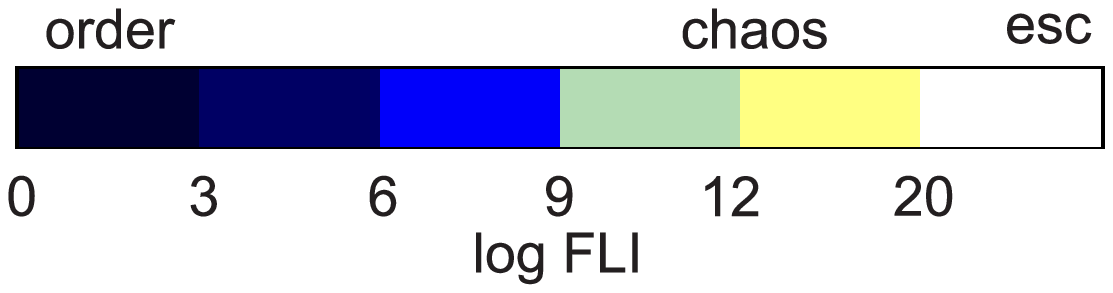} &  & \\
\textnormal{(c)} &  & 
\end{array} $
\caption{Stability maps of grid size $150\times 100$ along {\bf a.} family $f_e^{1/6}$ {\bf b.} family $f_e^{2/5}$. The color map is indicated in panel {\bf c} .}
\label{FigGridEHfam}
\end{figure}

In Figs. \ref{FigGridEHfam}a,b we show the stability maps for the families $f_e^{1/6}$ and $f_e^{2/5}$, respectively. Family $f_e^{1/6}$ is stable in the segments $-0.35<e_p<-0.23$ and $0.23<e_p<0.35$. The stability map shows the existence of two islands of stability (I1 and I2) around these segments. For $|e_p|>0.35$, where the family is unstable, strongly irregular orbits exist, which escape after a few periods. In the interval $-0.23<e_p<0.23$, where the family is also unstable, the existing instability gives a very thin chaotic zone in phase space, which is not clearly seen due to the resolution of the map. The wide stability regions (R1 and R2) that exist close to the unstable zone are associated with the stable family $f$ of the circular problem, as we have mentioned above. 

A similar dynamical situation is seen in the map of the family $f_e^{2/5}$. This family is stable for $-0.56<e_p<0$ and close to this segment we can observe the existence of regimes (I1-I3) with stable orbits. For $e_p>0$ the family is unstable, but for small eccentricities the instability zone is very thin. In this instability zone the orbits are weakly chaotic and their FLI evolution is like that of Fig. \ref{FigFLIevol} (case 3). The neighbouring stability regions R1 and R2 extend up to $e_p\approx 0.25$; for larger planetary eccentricities the orbits escape.   

\section{Conclusions}

The elliptic Hill's model is an extension of the famous circular Hill's model. It is obtained by applying the classical Hill's transformation to the equations of the elliptic restricted three body problem in a rotating frame. In this paper we studied the main qualitative features of its dynamics by computing (i) families of periodic orbits and their stability and (ii) stability maps for the determination of regions with regular orbits.

The families of periodic orbits in the EH model bifurcate from the families of the CH model and continue with parameter the planetary eccentricity, $e_p$. We can define an infinite set of bifurcation points on the CH families ($e_p=0$) by increasing the multiplicity of the periodic orbits. We have determined a large set of bifurcation points that correspond mainly to stable periodic orbits, since we are interested in locating stable families. For each bifurcation point we computed the family of periodic orbits by varying the planetary eccentricity, $e_p$. We classified the families according to their resonance and discussed their structure. We found that most of the families extend up to high absolute $e_p$ values for both the periapsis and the apoapsis case. However, in some cases the EH families reach a maximum value of $|e_p|$, where a saddle-node bifurcation takes place. Then a new EH family is generated, which also meet a bifurcation point on a CH family. Nevertheless, there may exist EH families that do not originate from critical points of the circular problem, but such families cannot be computed by the approximation followed in this paper.   

Most of the families are found to start as unstable, but in many cases family segments exist having stable periodic orbits even for relatively high absolute eccentricity values ($e_p>0.5$). Generally, continuing the families for $e_p>0$ and $e_p<0$ we find that at least one case gives simply unstable orbits. An exception is the family $f_e^{3/5}$, which in both cases starts as stable. Family segments of double and complex instability also exist.

In order to determine the phase space regions of regular orbits, we computed maps of dynamical stability defined on various planes. The majority of initial conditions corresponds to orbits that are irregular-chaotic and finally escape. However, the H\'enon stability maps showed that around the stable families $f$ and $g$ of the CH model there exist wide regions of regular motion. As the planetary eccentricity increases, the stability regions are affected significantly only when the planet is at apoapsis for $t=0$. Prograde orbits are affected more than retrograde ones.

The stable periodic orbits of the EH model are located at the ``centers'' of phase space regions with regular orbits. Instead, unstable periodic orbits are associated with chaotic orbits in their vicinity. For large absolute eccentricity values such chaotic orbits escape after a few planetary revolutions. For $|e_p|\lessapprox 0.2$ the chaotic regions around the unstable periodic orbits are thin and are surrounded by regions of regular orbits. In these regions the orbits are trapped and do not escape, at least for relatively long time spans.

\vspace{0.5cm}

\noindent{\bf Acknowledgements} The authors would like to thank the anonymous reviewers for their comments and fruitful suggestions to improve the
quality of the paper.

\end{document}